\documentclass[12pt, oneside]{article} 	
\usepackage{geometry}                		
\usepackage{graphicx}				
\usepackage{amssymb}
\usepackage{amsmath}
\DeclareMathOperator{\arccot}{arccot}
\usepackage{braket}
\usepackage{amsthm}
\usepackage{amsfonts}
\usepackage[colorlinks,linkcolor=blue,citecolor=blue,urlcolor=blue,
            bookmarks,bookmarksnumbered]{hyperref}
\usepackage[T1]{fontenc}
\usepackage[utf8]{inputenc}
\usepackage{authblk}

\usepackage{epsfig,epsf}
\input epsf.sty
\let\eps = \varepsilon
\def \be {\begin{equation}}
\def \ee {\end{equation}}
\def \bea {\begin{eqnarray}}
\def \eea {\end{eqnarray}}

\title{String field representation of the Virasoro algebra}

\author[1,2]{Nicholas Mertes\footnote{n.mertes@umiami.edu}}
\author[1]{Martin Schnabl\footnote{schnabl.martin@gmail.com}}

\affil[1]{\textit{Institute of Physics AS CR, Na Slovance 2, Prague 8, Czech Republic}}
\affil[2]{\textit{Department of Physics, University of Miami, Coral Gables, FL}}

\date{}

\begin{document}

\maketitle

\begin{abstract}
We construct a representation of the zero central charge Virasoro algebra using string fields in Witten's open bosonic string field theory. This construction is used to explore extensions of the $KBc$ algebra and find novel algebraic solutions of open string field theory.
\end{abstract}

\section{Introduction}
\label{sec-intro}

Our theory of interest is Witten's open bosonic string field theory~\cite{Witten86}. In particular, we focus on the research inspired by~\cite{Schnabl06}, where this theory was used to prove that the energy of the tachyon vacuum is minus the energy of the D-25 brane. There has been a good deal of progress made towards simplifying the original tachyon vacuum solution~\cite{Okawa, Erler09}. This simple solution is expressed in a language which makes use of an algebra of string fields called the $KBc$ algebra. Despite the advantages of this approach, there are various problems that can arise when using the $KBc$ algebra to construct solutions other than the tachyon vacuum~\cite{MS1,HK1,MS2,HK2}. Our attitude in this work is that these problems arise because the $KBc$ algebra is not large enough to properly describe all solutions of open string field theory.

If one tries to construct an algebra of string fields in an arbitrary way, the star products between the elements will be too difficult to manage. Therefore, in order to extend the $KBc$ algebra in a meaningful way, we must ask the question: Under what conditions can one straightforwardly construct an algebra of string fields with manageable star products? In the rest of this section we introduce one such family of states with simple star product structure.

Consider the operator modes $\phi_n$ corresponding to a primary conformal field $\phi(z)$ of weight $h$. These modes may be written as
\begin{equation}
\phi_n = \oint\frac{dz}{2\pi i}z^{n+h-1}\phi(z),
\end{equation}
where the contour of integration is along the complex unit circle unless otherwise specified. If we make a conformal transformation $\tilde{z} = f(z)$, then the transformed modes $\tilde{\phi}_n$ are given by
\begin{equation}
\tilde{\phi}_n = \oint\frac{dz}{2\pi i}\big[\partial f(z)\big]^{-h+1}[f(z)]^{n+h-1}\phi(z).
\end{equation}
It is convenient to introduce \mbox{$\tilde{v}_n(z) = \big[\partial f(z)\big]^{-h+1}[f(z)]^{n+h-1}$}, so we have
\begin{equation}
\tilde{\phi}_n = \oint\frac{dz}{2\pi i}\tilde{v}_n(z)\phi(z).
\end{equation}
The modes $\tilde{\phi}_n$ can be decomposed into their left and right components, respectively given by
\begin{equation}\label{tildeLRdef}
\begin{split}
\tilde{\phi}_n^L &= \int_L\frac{dz}{2\pi i}\tilde{v}_n(z)\phi(z), \\
\tilde{\phi}_n^R &= \int_R\frac{dz}{2\pi i}\tilde{v}_n(z)\phi(z),
\end{split}
\end{equation}
where $L$ denotes integration along the $\textrm{Re}(z)>0$ half of the complex unit circle and $R$ denotes integration along the $\textrm{Re}(z)<0$ half of the complex unit circle\footnote{This convention for defining left and right components is consistent with~\cite{Schnabl06, Erler09}. However, left and right components are defined in the opposite way in~\cite{Okawa}. The left-right decomposition of operators first appeared in the work~\cite{Horowitz:1986dta} and has been studied in many subsequent works. This splitting for the so called sliver frame $\tilde z = \arctan z$ has been of key importance for constructions of analytic solutions. More recently this construction has been revisited in the works~\cite{RZ,kz2,kz3}.}. Therefore, the sum of the left and right components corresponds to integration along the entire complex unit circle, namely, \mbox{$\tilde{\phi}_n^L + \tilde{\phi}_n^R =\tilde{\phi}_n$}. The left and right components of operators are interesting objects in string field theory because of their convenient properties with respect to the star product. For any string fields $A$ and $B$, we have
\begin{equation}\label{rule}
\begin{split}
(\tilde{\phi}_n^LA)\star B &= \tilde{\phi}_n^L(A\star B),\\
A\star(\tilde{\phi}_n^R B) &= (-1)^{|A| |\phi|} \tilde{\phi}_n^R(A\star B),
\end{split}
\end{equation}
where in the last line $|A|$ and $|\phi|$ refer to the ghost number of $A$ and $\phi$, respectively. We now focus our attention to the left components $\tilde{\phi}_n^L$. Let us define the string fields
\begin{equation}
\Phi^L_n = \tilde{\phi}_n^L\ket{I},
\end{equation}
where $\ket{I}$ is the identity of the star product. Using (\ref{rule}), we can compute the star commutator
\begin{equation}\label{lst}
\begin{split}
[\Phi^L_m,\Phi^L_n] &= \Phi^L_m\star \Phi^L_n - (-1)^{|\Phi^L_m||\Phi^L_n|}\Phi^L_n\star\Phi^L_m\\
&= \tilde{\phi}_m^L(\ket{I}\star \tilde{\phi}_n^L\ket{I}) - (-1)^{|\Phi^L_m||\Phi^L_n|}\tilde{\phi}_n^L(\ket{I}\star\tilde{\phi}_m^L\ket{I})\\
&= (\tilde{\phi}_m^L\tilde{\phi}_n^L - (-1)^{|\tilde{\phi}_m^L||\tilde{\phi}_n^L|} \tilde{\phi}_n^L\tilde{\phi}_m^L)\ket{I}\\
&= [\tilde{\phi}_m^L, \tilde{\phi}_n^L]\ket{I}.\\
\end{split}
\end{equation}
We note that in the above manipulation $|\Phi^L_n| = |\tilde{\phi}_n^L|$, since $\ket{I}$ has ghost number zero. Also, it will be assumed throughout that commutators are graded by ghost number. The result (\ref{lst}) shows us that if we create an algebra of string fields with elements $\Phi^L_n$, then the star commutators can be entirely determined by evaluating the \textit{ordinary} commutators \mbox{$[\tilde{\phi}_m^L, \tilde{\phi}_n^L]$}. We now provide an explanation of the conditions under which (\ref{lst}) can be written in terms of a
closed algebra of string fields.

Suppose that $\phi(z)$ belongs to a family of primary conformal fields indexed by a set $J$, $\phi(z) \in \{\phi^{\alpha}(z)\}_{\alpha\in J}$, such that
\begin{equation}
\label{phi-algebra}
[\phi_m,\phi_n] = \sum\limits_{\alpha\in J}f^\alpha_{mn}\,\phi^\alpha_{m+n},
\end{equation}
where the $f^{\alpha}_{mn}$ are constants and the modes $\phi^{\alpha}_n$ correspond to $\phi^{\alpha}(z)$. An important observation of this work is that there sometimes exists a conformal frame $\tilde{z}$ in which $\phi_n$ and $\tilde{\phi}_n^L$ satisfy the same algebra, meaning that
\begin{equation}\label{tildeLcom}
\begin{split}
[\tilde{\phi}^L_m,\tilde{\phi}^L_n] &= \sum\limits_{\alpha\in J}f^\alpha_{mn}(\tilde{\phi}^\alpha_{m+n})^L,
\end{split}
\end{equation}
where the constants $f^{\alpha}_{mn}$ are the same in both cases.\footnote{When the commutator $[\tilde{\phi}^L_m,\tilde{\phi}^L_n]$ is computed in an arbitrary conformal frame, it is common to encounter anomalous terms proportional to $\phi^{\alpha}(i)$ or its derivatives.} In this situation, it is straightforward to see that
\begin{equation}\label{7}
[\Phi^L_m,\Phi^L_n] = \sum\limits_{\alpha\in J}f^\alpha_{mn}(\Phi^\alpha_{m+n})^L.
\end{equation}
We refer to the algebra generated by (\ref{7}) as the \textit{string field representation} of the algebra (\ref{phi-algebra}) with structure constants $f^{\alpha}_{mn}$. This result answers the question posed at the beginning of this discussion. In particular, the condition (\ref{tildeLcom}) allows one to straightforwardly construct an algebra of string fields with manageable star \mbox{products}.~\footnote{Aside from the well known $K$ and $B$ string fields which form a trivial subalgebra of (\ref{7}), Erler \cite{Erler:2010pr} has noticed that a string field $G = {\cal G}_{-1/2}^L \ket{I} $ in the context of superstring field theory obeys the relation $G^2 = K$.}

The main observation of this work is that we have precisely the condition (\ref{tildeLcom}) for the matter plus ghost Virasoro modes $L_n$ and the conformal frame \mbox{$\tilde{z} = \frac{2}{\pi}\arctan{z}$}. The modes $L_n$ satisfy the zero central charge Virasoro algebra, which we now refer to simply as the Virasoro algebra. If we let \mbox{$\tilde{L}_n = \mathcal{L}_n$} and define the string fields \mbox{$\mathbb{L}_n = \mathcal{L}_n^L\ket{I}$}, then
\begin{equation}\label{strvir}
[\mathbb{L}_m, \mathbb{L}_n] = (m-n)\mathbb{L}_{m+n}.
\end{equation}
We call the algebra generated by (\ref{strvir}) the string field representation of the Virasoro algebra.

Naively, one would expect this discussion to be exactly the same for the right components. It is true that the modes $\mathcal{L}_n^R$ satisfy the Virasoro algebra, however, there is one interesting subtlety. Define the string fields ${\displaystyle \mathbb{L}^*_n} = \mathcal{L}^R_n\ket{I}$. Using the rule (\ref{rule}), we find
\begin{equation}\label{bpz}
[\mathbb{L}^*_m, \mathbb{L}^*_n] = -(m-n)\mathbb{L}^*_{m+n}.
\end{equation}
Recall that the BPZ dual of $L_n$ is given by $L_n^* = (-1)^nL_{-n}$. Therefore, the BPZ dual algebra is \mbox{$[L_m^*, L_n^*] = -(m-n)L^*_{m+n}$}. We see from this observation that the algebra generated by (\ref{bpz}) is actually the string field representation of the \textit{BPZ dual algebra}.

This paper is outlined as follows. In section \ref{sec-LR}, we prove the claim that $L_n$ together with the conformal frame \mbox{$\tilde{z} = \frac{2}{\pi}\arctan{z}$} satisfy the condition (\ref{tildeLcom}). We also discuss complications which arise if one considers correlation functions with operator insertions at the open string midpoint. In section \ref{sec-ExtKBc}, we discuss various options for extending the $KBc$ algebra. In section \ref{sec-app}, we use our results to find novel algebraic solutions of open string field theory. We conclude with a brief summary of our results and comment on opportunities for future exploration.

\section{Left and right components of $\mathcal{L}_n$}
\label{sec-LR}

Since we already know that \mbox{$[L_m,L_n] = (m-n)L_{m+n}$}, the main goal of this section is to show that \mbox{$[\mathcal{L}_m^L,\mathcal{L}_n^L] = (m-n)\mathcal{L}^L_{m+n}$}. The first step towards showing these commutation relations is to explicitly express the operator modes $\mathcal{L}_n$ as
\begin{equation}
\begin{split}
\mathcal{L}_n  &= \oint \frac{d\tilde{z}}{2\pi i}\tilde{z}^{n+1}\tilde{T}(\tilde{z}) \\
&= \bigg(\frac{2}{\pi}\bigg)^n\oint \frac{dz}{2\pi i} (1+ z^2)(\arctan{z})^{n+1}T(z).
\end{split}
\end{equation}
For the sake of brevity, we define
\begin{equation}
v_n(z) = \bigg(\frac{2}{\pi}\bigg)^n(1+ z^2)(\arctan{z})^{n+1}.
\end{equation}
An important property of the vector fields $v_n(z)$ is that
\begin{equation}\label{vanish}
\lim_{z\to \pm i}v_n(z) = 0.
\end{equation}
A quick but a bit formal argument for $[\mathcal{L}_m^L,\mathcal{L}_n^L] = (m-n)\mathcal{L}^L_{m+n}$ is to rewrite
\begin{equation}\label{ldef}
\mathcal{L}_n^L  =  \oint\frac{dz}{2\pi i}v_n(z)\theta(\textrm{Re}(z))T(z),
\end{equation}
where we have introduced the 'holomorphic step function'
\begin{equation}\label{step}
\theta(\textrm{Re}(z)) = \frac{1}{2} + \frac{1}{\pi}(\arctan{z} + \arccot{z})
\end{equation}
satisfying $\theta(\textrm{Re}(z)) = 1$ for $\textrm{Re}(z) > 0$ and $\theta(\textrm{Re}(z)) = 0$ for \mbox{$\textrm{Re}(z) < 0$}.

Let us define $L_v = \oint\frac{dz}{2\pi i} v(z) T(z)$.  Formally, or under suitable conditions, for any two such operators $ [L_v, L_w] = -L_{[v,w]}$, where
$[v,w] = v \partial w - w \partial v$ denotes the Lie bracket. Applying this formula for two vector fields of the form $v_n(z) \theta(\textrm{Re}(z))$ one quickly establishes
\be\label{LLcom}
[\mathcal{L}_m^L,\mathcal{L}_n^L] = (m-n)\mathcal{L}^L_{m+n}
\ee
thanks to the identity
\be
\label{vvid}
v_n(w) \partial v_m(w) - v_m(w) \partial v_n(w) = (m-n) v_{n+m} (w)
\ee
and the formal properties of the holomorphic step function $\theta(\textrm{Re}(z))^2=\theta(\textrm{Re}(z))$ and $\partial_z \theta(\textrm{Re}(z))=0 $, where especially the latter one may seem dubious.

\begin{figure}
\begin{center}
{\includegraphics[width={\textwidth}]{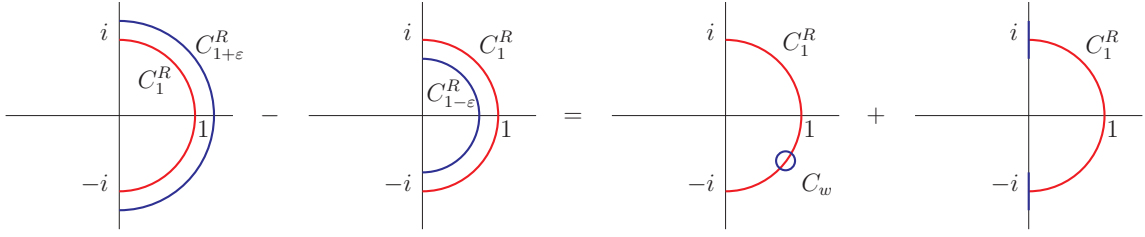}}
\end{center}
\caption{Contour manipulation involved in proving (\ref{LLcom}). The blue $z$-contour is deformed from $C_{1+\eps}^R$ into $C_{1-\eps}^R$ leaving behind small circle around the singularity at $w$, and integration along two open vertical segments of length $2\eps$. The variable $w$ is integrated along the unit red semicircle.}
\label{Fig-contours}
\end{figure}

Because of central importance of the equation (\ref{LLcom}) to this work, let us present a more rigorous derivation.
To proceed we follow the standard contour argument with modifications due to the step functions which restrict both contours to semicircles as in Figure \ref{Fig-contours}.
By picking sufficiently small $\eps>0$, the commutator can be expressed as
\be\label{contour_argument}
[\mathcal{L}_m^L,\mathcal{L}_n^L] = \left( \int_{C_{1+\eps}^R}
\int_{C_{1}^R} - \int_{C_{1-\eps}^R}
\int_{C_{1}^R}  \right)  \frac{dz}{2\pi i} \frac{dw}{2\pi i}  v_m(z) v_n(w) T(z) T(w).
\ee
In analogy to the standard argument we can deform the $z$-variable contour $C_{1+\eps}^R$ into the $C_{1-\eps}^R$ contour so that they cancel each other. What remains is a sum of two terms $I_1+I_2$

\bea
I_1 &=&  \int_{C_{1}^R} \frac{dw}{2\pi i} v_n(w) \oint_{C_w} \frac{dz}{2\pi i} v_m(z) T(z) T(w), \\
\label{I_2def}
I_2 &=&  \int_{C_{1}^R} \frac{dw}{2\pi i} v_n(w) \left(\int_{i-i\eps}^{i+i\eps} + \int_{-i-i \eps}^{-
i+i\eps}\right) \frac{dz}{2\pi i} v_m(z) T(z) T(w).
\eea
The first term $I_1$ is given by a small contour $C_w$ around the point $w$. Only the singular part of the zero-central-charge OPE
\be
T(z)T(w) \sim \frac{2T(w)}{(z-w)^2} + \frac{\partial T(w)}{z-w}
\ee
contributes. The resulting contribution takes the form
\bea
I_1 &=& \int_{C_{1}^R} \frac{dw}{2\pi i} v_n(w) \left[ 2 \partial v_m(w) T(w) +  v_m(w) \partial T(w) \right]
\nonumber\\
 &=& \int_{C_{1}^R} \frac{dw}{2\pi i} \left[ v_n(w) \partial v_m(w) - v_m(w) \partial v_n(w) \right] T(w)
\nonumber\\ && \quad +  v_n(i) v_m(i) T(i) +  v_n(-i) v_m(-i) T(-i),
\label{I_1res}
\eea
where in the second line we used integration by parts. The first term
already accounts for our desired commutation relation (\ref{LLcom}) thanks to (\ref{vvid}).
The last two terms are a contribution from the ends of the contour. They vanish since $v_n(\pm i) =0$, but this computation also hints at possible problems. Inserting the term $I_1$ inside a correlator with generic mid-point insertions at $\pm i$, the explicit appearance of $T(\pm i)$ on the right hand side of (\ref{I_1res}) would lead to divergences and the final result would depend on the precise manner the various limits are taken. In this work we allow the probing operator to approach the midpoint at the very end, so that the midpoint contributions in (\ref{I_1res}) are effectively absent.

The second term $I_2$ is naively not contributing either, since the
integration region is of vanishing size as $\eps$ is taken to zero. One must exercise a caution however, since the integrand of the double integral (\ref{I_2def}) is divergent in the region $z \sim w \sim \pm i$ where its leading singularity is given by
\be
\frac{ t u}{(i t - u)^2} (\log t)^{m+1}  (\log u)^{n+1}  T(u).
\ee
Here we parameterized $z = \pm i + i t$ and $w = \pm i + u$.
Standard power-counting argument---assuming regular $T(u)$---shows however that the singularity is integrable, and hence the contribution from a vanishing size region is indeed vanishing. The easiest way to see this is by passing to radial coordinates in the $t$-$u$
plane. On the other hand, had we evaluated $I_2$ inside a correlator with midpoint insertions, we would have to replace $T(u)$ by e.g. $1/u^2$  in case of a primary field, and  $I_2$ would have been divergent, or at least prescription dependent.

Let us close this section by noting that similarly to (\ref{ldef}), we have
\begin{equation}
\mathcal{L}_n^R = \oint\frac{dz}{2\pi i}v_n(z)\theta(\textrm{Re}(-z))T(z),
\end{equation}
where $\theta(\textrm{Re}(-z))$ corresponds to integration along the
$\textrm{Re}(z)<0$ half of the complex unit circle. We find the expected result
\begin{equation}
[\mathcal{L}_m^R,\mathcal{L}_n^R] = (m-n)\mathcal{L}_{m+n}^R.
\end{equation}
One may also consider commutators involving both left and right components. Using a similar analysis, we obtain\footnote{This commutation relation has a controversial history. The result given here is consistent with~\cite{RZ}, but inconsistent with~\cite{kz2, kz3}. The issue is that this commutation relation breaks down when one considers correlation functions with operator insertions at the open string midpoint $z = i$. For the sake of the current paper, we do not consider such situations. However, this issue is a current topic of interest and will be addressed in~\cite{msmh}.}
\begin{equation}
[\mathcal{L}_m^L,\mathcal{L}_n^R] = 0.
\end{equation}

\section{Extensions of the $KBc$ algebra}
\label{sec-ExtKBc}

We first recall the most important results of sections \ref{sec-intro} and \ref{sec-LR}. We have shown that the strings fields \mbox{$\mathbb{L}_n = \mathcal{L}^L_n\ket{I}$} form a representation of the Virasoro algebra
\begin{equation}
[\mathbb{L}_m, \mathbb{L}_n] = (m-n)\mathbb{L}_{m+n}.
\end{equation}
We have also seen that the string fields \mbox{$\mathbb{L}_n^* = \mathcal{L}^R_n\ket{I}$} form a representation of the BPZ dual algebra
\begin{equation}
[\mathbb{L}^*_m, \mathbb{L}^*_n] = -(m-n)\mathbb{L}^*_{m+n}.
\end{equation}
At this point we restrict our discussion
for convenience to the Virasoro algebra rather than its dual.
However, similar remarks will hold if one wishes to formulate a theory using the BPZ dual algebra.

The string fields of the $KBc$ algebra are defined in our language as
\begin{align}
K = \mathcal{L}_{-1}^L\ket{I} = \mathbb{L}_{-1}, && B = \tilde{b}_{-1}^L\ket{I}, && c = \tilde{c}\bigg(\frac{1}{2}\bigg)\ket{I},
\end{align}
where as before the tilde refers to the operator expressed in the \mbox{$\tilde{z} = \frac{2}{\pi}\arctan{z}$} conformal frame. Given our previous observations, it is clear that we may define an algebra of anticommuting string fields analogous to $B$, given by
\begin{equation}
\mathbb{B}_n = \tilde{b}^L_n\ket{I},
\end{equation}
so that $B = \mathbb{B}_{-1}$. If we use the convention that all star commutators are graded by ghost number, then the standard algebraic properties of $KBc$ are
\begin{align}\label{old}
[\mathbb{L}_{-1}, \mathbb{B}_{-1}] &= 0, & Q\mathbb{L}_{-1}  &= 0, \nonumber\\
[\mathbb{L}_{-1}, c] &= \partial c, & Q\mathbb{B}_{-1} &= \mathbb{L}_{-1},\\
[\mathbb{B}_{-1}, c] &= 1, & Qc &= c\,\mathbb{L}_{-1} c, \nonumber
\end{align}
where $\partial c = \tilde{\partial}\tilde{c}\big(\frac{1}{2}\big)\ket{I}$.

We are now prepared to begin our discussion of extending the $KBc$ algebra. The most straightforward extension is to include all of the string fields $\mathbb{L}_{n}$ and $\mathbb{B}_{n}$. This algebra satisfies
\begin{align}\label{newbig}
[\mathbb{L}_{m}, \mathbb{B}_{n}] &= (m - n)\mathbb{B}_{m+n}, & Q\mathbb{L}_{n}  &= 0,\\
[\mathbb{B}_{m}, c] &= \bigg(\frac{1}{2}\bigg)^{m+1}, & Q\mathbb{B}_{n} &= \mathbb{L}_{n}.\nonumber
\end{align}
The commutator \mbox{$[\mathbb{L}_{n}, c]$} is a bit trickier to evaluate. We begin by computing
\begin{equation}
\begin{split}
\mathbb{L}_n\star c &= \mathcal{L}_n^L c, \\
c\star \mathbb{L}_n &= -\mathcal{L}_n^Rc + c\star(\mathcal{L}_n\ket{I}),\\
[\mathbb{L}_n, c] &= \mathcal{L}_n c - c\star(\mathcal{L}_n\ket{I}).
\end{split}
\end{equation}
Using the rule \mbox{$(\tilde{c}\big(\frac{1}{2}\big)\ket{I})\star(\mathcal{L}_n\ket{I})= \tilde{c}\big(\frac{1}{2}\big)\mathcal{L}_n\ket{I}$}, we see that
\begin{equation}\label{Lc-com}
\begin{split}
[\mathbb{L}_n, c] &= \bigg[\mathcal{L}_n, \tilde{c}\bigg(\frac{1}{2}\bigg)\bigg]\ket{I}.
\end{split}
\end{equation}
By computing the commutator (\ref{Lc-com}), we find
\be
[\mathbb{L}_{n}, c] =[-(n+1)(2)^{-n} + (2)^{-n-1}\tilde{\partial}]\tilde{c}\bigg(\frac{1}{2}\bigg)\ket{I}
\ee
and hence
\be
Qc = (2)^{n+1}c\,\mathbb{L}_n c.
\ee
Notice that by defining $\mathbb{C}_n = (2)^{n+1}c$, we have
\begin{equation}\label{spcon}
\begin{split}
Q\mathbb{C}_n &= \mathbb{C}_n\mathbb{L}_n\mathbb{C}_n,\\
[\mathbb{B}_{m}, \mathbb{C}_n] &= (2)^{n-m}.
\end{split}
\end{equation}
The above relationships will be convenient when we use these results to write down solutions of open string field theory.

Another possibility is to consider a finite extension of the $KBc$ algebra. In fact, a finite extension is often computationally easier to understand. The infinite extension was introduced first in order to develop all of the techniques necessary to deal with extended $KBc$ algebras. The Virasoro algebra has two non-trivial finite subalgebras.
The elements $\mathbb{L}_{-1}$ and $\mathbb{L}_{0}$ form what is called $\mathfrak{aff}(1)$, while  $\mathbb{L}_{-1}$, $\mathbb{L}_{0}$, and $\mathbb{L}_{1}$ form the familiar $SL(2,R)$ algebra.
By adding the $\mathbb{B}$ counterparts to either $\mathfrak{aff}(1)$ or $SL(2,R)$, one can create a finite extension
of the $KBc$ algebra. However, it should be noted that one is not strictly limited to these cases. For the generating string fields
one could instead choose various linear combinations of the elements in $\mathfrak{aff}(1)$ or $SL(2,R)$. In fact, one of the early motivations for this work was the discovery of one of the $\mathfrak{aff}(1)$ linear combinations. Consider the string field
\begin{equation}
L = \mathbb{L}_{0} - \frac{1}{2}\mathbb{L}_{-1}.
\end{equation}
The string field $L$ has the interesting property that
\begin{equation}\label{lminus}
\begin{split}
[L,K] &= L^-K = K,\\
[L,B] &=L^-B = B,\\
[L,c] &=L^-c = -c,
\end{split}
\end{equation}
where \mbox{$L^- = \frac{1}{2}(\mathcal{L}_0 - \mathcal{L}_0^*)$} is the familiar derivation of the star algebra. Therefore, adding $L$ to the $KBc$ algebra is effectively turning an exterior derivative into an interior derivative.

The $SL(2,R)$ extension also has a nice physical interpretation. It is well-known that the string field $K$ acts as a generator of translations when acting on the string field $c$, a fact which is expressed by the commutation relation $[K,c] = \partial c$. It turns out that the $SL(2,R)$ extension effectively adds the generators of dilation and special conformal transformation to the $KBc$ algebra. To see this, let us use the convention that the generator $\mathcal{G}_n$ associated with $\mathbb{L}_n$ is defined implicitly by \mbox{$[\mathbb{L}_{n},c] = i\mathcal{G}_nc$}. Then we see from (\ref{spcon}) that
\begin{equation}
\mathcal{G}_n = -i\big[-(n+1)\tilde{z}^n + \tilde{z}^{n+1}\tilde{\partial}\big],
\end{equation}
where we let $\tilde{z}\to \frac{1}{2}$. Notice that the generators $\mathcal{G}_{-1}$, $\mathcal{G}_{0}$, and $\mathcal{G}_{1}$ form a representation of the holomorphic component of the global conformal algebra in two dimensions.

\section{Application to algebraic solutions}
\label{sec-app}

The goal of this section is to show that we can use the extensions described in section \ref{sec-ExtKBc} to construct new analytic solutions to the open string field theory equation of motion \mbox{$Q\Psi + \Psi\star\Psi = 0$}. We begin by reviewing important techniques for constructing solutions with the $KBc$ algebra. The modern method of constructing solutions is to start with the pure gauge ansatz \mbox{$U = 1 - FBcF$}, where $F = F(K)$ is an appropriate function~\cite{Schnabl:2010tb} of the string field $K$. The corresponding solution $\Psi$ is given by $\Psi = UQU^{-1}$. In order to find $U^{-1}$, we compute
\begin{equation}
\begin{split}
U^{-1} &= 1 + \sum\limits_{n = 1}^\infty (FBcF)^{n-1}FBcF\\
&= 1 + \sum\limits_{n = 1}^\infty (F^2)^{n-1}FBcF\\
&= 1 + \frac{1}{1-F^2}FBcF.
\end{split}
\end{equation}
This result for $U^{-1}$ leads to the solution
\begin{equation}\label{15}
\Psi = FcB\frac{1}{1-F^2}KcF,
\end{equation}
which is the familiar Okawa ansatz discovered in~\cite{Okawa}.

In the spirit of section \ref{sec-ExtKBc}, we first show how this solution nicely generalizes to the infinite extension of $KBc$, which includes all of the string fields $\mathbb{L}_n$, $\mathbb{B}_n$, and $\mathbb{C}_n$. We start with the pure gauge ansatz
\begin{equation}
U_n = F(\mathbb{L}_n)\mathbb{B}_n\mathbb{C}_nF(\mathbb{L}_n)
\end{equation}
for a fixed choice of $n$.
The relationships (\ref{newbig}) and (\ref{spcon}) tell us that for each fixed value of $n$, the
$\mathbb{L}_n, \mathbb{B}_n, \mathbb{C}_n$ algebra
behaves almost identically to the original $KBc$ algebra.  Therefore, we find that the pure gauge ansatz $U_n$ leads to the solution
\begin{equation}\label{bigsol}
\begin{split}
\Psi_n=F(\mathbb{L}_n)\mathbb{C}_n\mathbb{B}_n\frac{1}{1-F^2(\mathbb{L}_n)}\mathbb{L}_n\mathbb{C}_nF(\mathbb{L}_n).
\end{split}
\end{equation}

The computation and classification of all possible solutions using different representations of the $KBc$ algebra
is certainly a project in itself, and therefore we do not attempt to perform such an analysis here.\footnote{It would be interesting for instance to compare this solution to the ones constructed in \cite{RZ, ORZ}.} However, we conclude this work by drawing attention to some interesting solutions that are of a different form than (\ref{bigsol}). These solutions are perhaps best illustrated by focusing on an algebra similar to the one described in (\ref{lminus}). This algebra is defined by extending the $KBc$ algebra with the two string fields
\begin{equation}
\begin{split}
L &= \mathbb{L}_{0} - \frac{1}{2}\mathbb{L}_{-1},\\
B' &= \mathbb{B}_{0} - \frac{1}{2}\mathbb{B}_{-1}.
\end{split}
\end{equation}
In addition to the usual $KBc$ commutation relations, the $KLBB'c$ algebra satisfies
\begin{align}
[L,K] &= K, & [B',K] &= B, & [L,B] &= B, \nonumber\\
[L, B'] &= 0, & [L,c] &= -c, & [B,B'] &= 0,\\
[B',c] &= 0, & QB' &= L, & QL &= 0. \nonumber
\end{align}

Analytic solutions with the $KLBB'c$ algebra can involve functions of the type $F = F(K,L)$. It is important to note that this is a function of non-commuting variables. Therefore, some operations on $F(K,L)$ will depend on how the function is ordered. In effort to state our results in the most convenient way, let us adopt the convention that we always order $F(K,L)$ with every $K$ appearing to the left of every $L$. In particular, the functions we will be interested in are those which can be written as
\begin{equation}\label{fkl}
F(K,L) = \int d\alpha \int d\beta\:f(\alpha, \beta)\,e^{\alpha K}e^{\beta L}.
\end{equation}
First note that \mbox{$BL^m = (L-1)^m B$}. Therefore, since $B$ commutes with $K$, we have\footnote{We thank Ondra Hul\'{i}k for helping us to better understand manipulations involving the $KLBB'c$ algebra.}
\begin{equation}\label{min}
BF(K,L) = F(K,L-1)B.
\end{equation}
It is similarly true that \mbox{$KF(K,L) = F(K,L-1)K$}. Next we study the way $B'$ interacts with $F(K,L)$. Notice that since \mbox{$[[B',K],K] = 0$}, we can use the formula
\begin{equation}
[B',K^n] = nK^{n-1}[B',K] = nK^{n-1}B.
\end{equation}
Recalling the form of our function (\ref{fkl}) and the rule (\ref{min}), we find that $B'$ satisfies the commutation relation
\begin{equation}\label{par}
[B',F(K,L)] = \frac{\partial{F(K,L-1)}}{\partial K}B.
\end{equation}
The above equation raises an additional concern. Even with our prescribed ordering of the function $F(K,L)$, there is still some confusion that may arise when taking the $K$ derivative of a product of functions. Consider two functions $A(K,L)$ and $B(K,L)$ of the type (\ref{fkl}). The correct product rule is
\begin{equation}\label{prod}
\frac{\partial}{\partial K}\big[A(K,L)B(K,L)\big] = \frac{\partial A(K,L)}{\partial K}B(K,L) + A(K,L+1)\frac{\partial B(K,L)}{\partial K}.
\end{equation}
The above equation is actually quite intuitively pleasing. Since passing $K$ through $A(K,L)$ sends \mbox{$L\to L-1$}, it makes sense that passing $\frac{\partial}{\partial K}$ through $A(K,L)$ sends \mbox{$L \to L + 1$}.

Recall the pure gauge ansatz \mbox{$U = 1 - FBcF$}. We obtain a new type of analytic solution by changing \mbox{$F(K)\to F(K,L)$}. Using the relationship (\ref{min}), we find
\begin{equation}
\Psi = F(K,L)cB\frac{1}{1-F^2(K,L)}KcF(K,L).
\end{equation}
Notice, that the noncommutativity of $K$ and $L$ requires positioning the $K$ and $B$ factors appropriately on the two sides of the middle factor $(1-F^2(K,L))^{-1}$. We postpone the detailed analysis of the physics of such solutions to a future work.

Notice that we also have the freedom to consider a more general $U$ of the form
\begin{equation}
U = 1 - \sum\limits_i F_i^LBcF_i^R,
\end{equation}
where $F_i^L$ and $F_i^R$ are both functions of the string fields $K$ and $L$. The inverse of this pure gauge choice is given by
\begin{equation}
\begin{split}
U^{-1} &= 1 + \sum\limits_{n=1}^\infty\bigg(\sum\limits_i F_i^LBcF_i^R\bigg)^{n-1}\sum\limits_j F_j^LBcF_j^R\\
&= 1 + \sum\limits_{n=1}^\infty\bigg(\sum\limits_i F_i^LF_i^R\bigg)^{n-1}\sum\limits_j F_j^LBcF_j^R\\
&= 1 + \frac{1}{1-\sum\limits_i F_i^LF_i^R}\:\sum\limits_j F_j^LBcF_j^R.
\end{split}
\end{equation}
The solution $\Psi = UQU^{-1}$ looks somewhat more complicated because the summation prohibits some of our usual simplification tricks. However, it can still be written in the reasonably convenient form\footnote{Similar solutions have been explored in~\cite{MNT}.}
\begin{equation}
\Psi = Q\bigg(\sum\limits_i F_i^LBcF_i^R\bigg) + \sum\limits_i F_i^LcBF_i^R\frac{1}{1-\sum\limits_j F_j^LF_j^R}\sum\limits_k F_k^LKcF_k^R.
\end{equation}
This is a convenient form because the term $Q(\dots)$ will not contribute to a computation of the energy.

By incorporating the new string field $B'$, we can consider a pure gauge ansatz \mbox{$U = 1 - F(K,L)B'cF(K,L)$}. To efficiently compute $U^{-1}$, recall (\ref{par}) and notice that
\begin{equation}
\big(B'cF^2(K,L)\big)B' = \bigg(-c\frac{\partial}{\partial K}\big(F^2(K,L-1)\big)B\bigg)B'.
\end{equation}
Now we can compute
\begin{equation}
\begin{split}
U^{-1} &= 1 + \sum\limits_{n=1}^{\infty}(F(K,L)B'cF(K,L))^n\\
&= 1 + F(K,L)\sum\limits_{n=1}^{\infty}\bigg(-c\frac{\partial}{\partial K}\big(F^2(K,L-1)\big)B\bigg)^{n-1}B'cF(K,L)\\
&= 1 + F(K,L)c\sum\limits_{n=1}^{\infty}\bigg(-\frac{\partial}{\partial K}\big(F^2(K,L-1)\big)\bigg)^{n-1}BB'cF(K,L)\\
&= 1 + F(K,L)cB\frac{1}{1 + \frac{\partial}{\partial K}(F^2(K,L))}B'cF(K,L).
\end{split}
\end{equation}
As usual, the corresponding solution is given by $\Psi = UQU^{-1}$. When dealing with solutions of the above type, one must be careful to use the rule (\ref{prod}) when computing the $K$ derivative of a product of functions.

One motivation for writing down these new solutions is in effort
to construct solutions describing multiple D-brane configurations. Such an attempt has already been made in~\cite{MS1}, but the solutions unfortunately failed some tests of regularity \cite{HK1,MS2,HK2}. It is our hope that the extended $KBc$ framework can be used to write down regular multibrane solutions.  Also, it is apparent that some of these results can be extended to superstring field theory, following the work of~\cite{Erler:2010pr, AldoArroyo:2012if}.

\section{Conclusion}
We have constructed an interesting new representation of the zero central charge Virasoro algebra using string fields. We have used this construction to explore various extensions of the $KBc$ algebra, and find new analytic solutions of open string field theory. It should be noted that many proposals have been made in the past which attempt to make use of left and right components. Some of these proposals are~\cite{HM, gt1, gt2, Er1, Er2}. However, our work comes from a quite different perspective. We have been able to make insights not easily seen in previous constructions. The observations of this work are rich with opportunities for future exploration. Such exploration would include studying the surfaces
generated by $\mathbb{L}_n$, computing the energy of the new solutions, and writing down a most general solution to the open string field theory equation of motion. This might necessitate revisiting and generalizing previous results, especially elucidating the role played by the so called hidden boundary \cite{kz2,kz3}.

\section*{Acknowledgements}
N.M. thanks the Institute of Physics AS CR for their hospitality during the preparation of this work. N.M. also thanks M.S. for introducing him to the subject of algebraic solutions in open string field theory. The research of M.S. has been supported by the Grant Agency of the Czech Republic, under the grant 14-31689S. N.M. and M.S. thank Ondra Hul\'{\i}k and Carlo Maccaferri for useful discussions regarding this work.

\end{document}